\documentclass[twocolumn,superscriptaddress,showpacs]{revtex4}
\newcommand{\one}{\mbox{$1 \hspace{-1.0mm}  {\bf l}$}}

\begin{document}

\title{Entangled states maximize the
two qubit channel capacity for some Pauli channels with memory.}
\author{Chiara Macchiavello}
\affiliation{Dipartimento di Fisica ``A.Volta", Via Bassi 6,
I-27100 Pavia, Italy,
and Istituto Nazionale per la Fisica della Materia (INFM)}
\author{G.Massimo Palma}
\affiliation{INFM - NEST\& Dipartimento di Tecnologie
dell'Informazione, Universita' degli studi di Milano via Bramante
65, I-26013 Crema (CR), Italy}
\author{Shashank Virmani}
\affiliation{QOLS, Department of Physics, Blackett Laboratory,
Imperial College, Prince Consort Road, London SW7 2BW, UK}

\date{\today}
\begin{abstract}
We prove that a general upper bound on the maximal mutual
information of quantum channels is saturated in the case of Pauli
channels with an arbitrary degree of memory. For a subset of such
channels we explicitly identify the optimal signal states. We show
analytically that for such a class of channels entangled states
are indeed optimal above a given memory threshold. It is
noteworthy that the resulting channel capacity is a
non-differentiable function of the memory parameter.
\end{abstract}


\maketitle

%%%%%%%%%%%%%%%%%%%%%%%%%%%%%%%%%%%%%%%%%%%%%%%%%%%%%%%%%%%%%%%%%%%%%%%%%%%%%

The study of the optimal coding and decoding of information in
quantum systems has a long history \cite{Helstrom}. The advent of
present day quantum information theory \cite{Nielsen} \cite{WS}
has not only revived interest in the subject but has also opened
new problems. A key open question concerns the additivity of
channel capacity when entangled states are used as signals.
Although entanglement is a ubiquitous ingredient in nearly all
quantum information processing protocols and algorithms, it is
often regarded as being very fragile in the presence of
environmental noise. This has led to the belief that in most
circumstances the use of entanglement is not advantageous in the
reliable transmission of classical information through quantum
channels. For those memoryless channels (i.e. ones in which the
noise acting on consecutive uses of the channel is uncorrelated)
that have been studied so far, this is indeed the case. This was
first proven analytically for the depolarizing channel
\cite{depol}, where isotropic noise acts on individual qubits, and
then extended to a more general form of memoryless unital channel
\cite{King}. There has also been interesting recent work
demonstrating that the potential additivity of channel capacities
is equivalent to other well known additivity conjectures in
quantum information theory \cite{shor}. The scenario changes when
the channel is not memoryless, i.e. when the noise acting on
consecutive uses is partially correlated. This phenomenon is not
uncommon in physical situations, when the statistical properties
of the physical source of noise can be time - dependent. The
problem of quantum channels with memory was first introduced in
\cite{MP}, where, for the case of depolarizing channels with
memory, it was shown that the use of entangled states enhances the
mutual information. In Ref. \cite{MP} input states taken from a
certain ansatz were considered, and it was shown that within this
ansatz entangled states allow for the transmission of a larger
amount of reliable information. However, it was not proved
analytically that this ansatz is indeed optimal. Further results
bounding the asymptotic capacities of noisy channels with memory
have also recently been derived \cite{bm}.

Here for the first time we prove the optimality of a set of entangled input signal states for a class
of Pauli channels. To this end we will first obtain an upper bound on the channel capacity. We will then show
that for the general case of Pauli channels with an arbitrary degree of memory this bound is saturated by states
of minimal output entropy. For a class of Pauli channels we will derive these states explicitly. They turn out
to be entangled above a given memory threshold and product states below it.

In order to set the scenario let us first consider a single qubit
channel that is a random implementation of the Pauli
transformations:

\begin{equation}
\rho  \rightarrow  \sum^{3}_{i=0}  q_{i}  \sigma_{i}  \rho
\sigma_{i}. \label{single}
\end{equation}
where the $q_i$ give a probability distribution, and the
$\sigma_i$ are the Pauli matrices according to the following
convention:
\begin{eqnarray}
  \sigma_0 = \left(\begin{array}{cc}
    1  & 0 \\
     0 & 1 \end{array}\right)~;~  \sigma_1 = \left(\begin{array}{cc}
    1  & 0 \\
     0 & -1 \end{array}\right) \nonumber \\
      \sigma_2 = \left(\begin{array}{cc}
     0 & 1 \\
     1 & 0 \end{array}\right)~;~  \sigma_3 = \left(\begin{array}{cc}
    0  & -i \\
     i & 0 \end{array}\right).
\end{eqnarray}
We will sometimes refer to the Pauli matrices as the Pauli group,
even though extra phases are required in order to make the matrices
closed under matrix multiplication. However, since these phases
cancel out when considering transformations of density operators,
as
\begin{equation}
(e^{i\theta}\sigma_i) \rho (e^{i\theta}\sigma_i)^{\dag} = \sigma_i
\rho \sigma_i,
\end{equation}
we will freely make this abuse of terminology.

In the typical memoryless channel scenario, an understanding of
the action of an individual use, such as the one described in
Eq.(\ref{single}), is sufficient to fully describe the operation
of the channel. However, the possibility of repeated access to the
channel opens the question of optimizing the choice of signal
states, including the ones that are entangled over many uses of
the channel. This leads to questions concerning the additivity of
channel capacities, and whether entangled inputs and output
measurements can lead to improvements in information transmission.
However, in the manner of \cite{MP} we would like to consider
repeated applications of a single qubit channel that are {\it not}
independent. In particular we will consider a two qubit channel
that is almost equivalent to two independent uses of the single
qubit channel (\ref{single}), aside from a memory factor $ \mu \in
[0,1]$ that introduces correlations, i.e.:

\begin{equation}
\rho  \rightarrow  \sum^{3}_{i,j=0}  p_{ij}  \sigma_{i} \otimes
\sigma_{j}  \rho  \sigma_{i}  \otimes  \sigma_{j} \nonumber
\end{equation}
where
\begin{equation}
p_{ij} = (1-\mu) q_{i} q_{j} + \mu q_{i} \delta_{ij}.
\label{memory}
\end{equation}

We can see that this evolution can be considered as two independent applications of (\ref{single}), except for
an additional effect due to the degree of memory $\mu$, which with some probability forces the same Pauli
transformation to be repeated in the second use of the channel.

We would like to compute the maximum amount of information that
can be transmitted through a noisy channel of the form
(\ref{memory}), and investigate how the use of entangled inputs in
the two uses of the channel may improve its communication
performance. To do this we will show that this is equivalent to
finding the input pure state with minimal output entropy.

The maximum mutual information of a general quantum channel
${\cal{E}}$ is given by
the Holevo-Schumacher-Westmoreland bound \cite{HSW}:
\begin{equation}
\chi ({\cal{E}}) = \max_{\{p_i, \rho_i\}} S({\cal{E}}(\sum_{i}
p_{i} \rho_{i})) - \sum_i p_i S({\cal{E}}(\rho_i)) \label{HSW}
\end{equation}
where $S(\omega ) = - {\mbox{Tr}}(\omega \log \omega )$ is the von
Neumann entropy of the density operator $\omega$ and the
maximization is performed over all input ensembles $\{p_i,
\rho_i\}$ into the channel ($\rho_i$ are the input states on which
classical information is encoded, and are transmitted with prior
probabilities $p_i$). Note that this bound incorporates a
maximization over all POVM measurements at the receiver, including
collective ones over multiple uses of the channel.

In our scenario the $\rho_i$ describe states of two qubits, and so
we will refer to the maximum mutual information $\chi ({\cal{E}})$
as the {\it two-qubit} capacity of the channel. We will find it
convenient to use the symbol $\rho_{*}({\cal{E}})$ to denote a
chosen input state that gives minimal output entropy when
transmitted through the channel ${\cal{E}}$. As the maximally
mixed state gives the largest possible entropy for any system, the
formula (\ref{HSW}) can clearly be bounded from above by
\begin{equation}
\chi ({\cal{E}}) \leq \log _2(4) - S(\rho_{*}({\cal{E}})) = 2 -
S(\rho_{*}({\cal{E}})). \label{bound}
\end{equation}
for any 2-qubit channel. We will now see that this upper bound can be achieved by any 2-qubit channel whose
action consists of random tensor products of Pauli transformations.
The argument that we use to
demonstrate this can be applied to any channel that is covariant with respect to an irreducible representation
of a compact group, and has been independently noted by Holevo \cite{hol}.
The key ingredients will
be the facts that the Pauli matrices (a) form an irreducible representation of a group, and (b) either commute
or anticommute. Indeed, as these are essentially the only ingredients required, the same argument can easily be
modified to multiqubit channels whose actions consist of random tensor products of Pauli matrices.

Let us consider an ensemble of input states given by the sixteen
states defined by $\rho_{ij} := \sigma_i \otimes \sigma_j \rho_{*}
\sigma_i \otimes \sigma_j $, each with the same input probability
1/16. The commutation relations of the Pauli matrices imply that
any channel ${\cal{E}}$ of the form (\ref{memory}) is covariant
with respect to the Pauli rotations
\begin{equation}
 {\cal{E}} (\sigma_i \otimes \sigma_j \rho_{*} \sigma_i \otimes
 \sigma_j) = \sigma_i \otimes \sigma_j {\cal{E}}(\rho_{*}) \sigma_i \otimes
\sigma_j
\end{equation}
As entropy is invariant under unitary transformations, we can
immediately write
\begin{equation}
S({\cal{E}}(\rho_{*})) = S({\cal{E}}(\rho_{ij})), \label{one}
\end{equation}
and therefore each of the states $\rho_{ij}$ will also give the
same minimal output entropy as $\rho_{*}$. Furthermore, the fact
that the group of matrices $\{\sigma_i \otimes \sigma_j\}$ is an
irreducible representation means that the ensemble will give an
average output state that is maximally mixed \cite{ir}
\begin{equation}
{\cal{E}}\left(\sum_{ij} \frac{1}{16} \rho_{ij}\right) = \sum_{ij}
{1 \over 16} \sigma_{i} \otimes  \sigma_{j} {\cal{E}}(\rho)
\sigma_{i} \otimes \sigma_{j} =\frac{\one}{4}\;.
\label{two}
\end{equation}
Inserting equations (\ref{one}) and (\ref{two}) into equation
(\ref{HSW}) we can see that the upper bound (\ref{bound}) is attained
by the input ensemble of states $\rho_{ij} := \sigma_i \otimes
\sigma_j \rho_{*} \sigma_i \otimes \sigma_j$ with equal prior
probabilities. This means that to optimise the information transmission
of our channel, we merely need to search for the input state that
minimises the output entropy. We will refer to any such state as
an {\it optimal} input state.

In \cite{MP} a specific form of memory
channel was investigated, where the weights in equation (\ref{memory}) were
fixed by
\begin{equation}
q_0 = x~~~~~~;~~~~~~q_1 = q_2 = q_3 = {1-x \over 3}
\end{equation}
and the degree of memory $\mu$ was allowed to take any value in
the interval $[0,1]$. An ansatz for the form of the optimal input
state was conjectured, but a full analytic proof is still lacking.
%Later we will discuss further numerical evidence
%for the validity of their ansatz for channels with more general
%values for $q_i$, and present further numerical evidence for its
%validity.

Consequently, here we will focus our attention on a kind of memory channel
for which we can give an entirely analytic solution. The form of the channel is
characterized by the following parameters in equation (\ref{memory})
\begin{equation}
q_0 = q_1 = p ~~~; ~~~ q_2 = q_3 = q\;,
\end{equation}
where $q=(1-2p)/2$.

%This channel, including memory, is a particular probabilistic
%mixture of unitary transformations of the form $\sigma_i \otimes
%\sigma_j$ with weights $p_{ij}$. We can therefore describe the
%action of the channel with the matrix consisting of elements
%$p_{ij}$
%\begin{equation}
%  { \cal{E} }= (1-\mu) \left(\begin{array}{cccc}
%    p^{2} & p^{2} & p.q & p.q\\
%    p^{2} & p^{2} & p.q & p.q \\
%    p.q & p.q & q^2 & q^2 \\
%    p.q & p.q & q^{2} & q^2 \end{array}\right)
%     +
%     \mu  \left(\begin{array}{cccc}
%    p & 0 & 0 &  0 \\
%    0 & p & 0 &  0 \\
%    0 & 0 & q & 0 \\
%    0 & 0 & 0 & q \end{array}\right)
%\end{equation}
In order to identify the optimal input states we will first show
that we can restrict our attention to input states that are
invariant under the symmetry group $\{\sigma_0 \otimes \sigma_0,
\sigma_1 \otimes \sigma_1\}$. The technique that we will use may
be generalised to many other channels with a suitable structure
\cite{general}. Let us first consider the following modification
of the channel $\cal{E}$: first rotate the input state by
$\sigma_1 \otimes \sigma_1$, and then act with $\cal{E}$. Let us
call this new channel $\cal{E}'$:= ${\cal{E}} \circ (\sigma_1
\otimes \sigma_1) $. Using the standard relations for the Pauli
group: $\sigma_0 \sigma_1 = \sigma_1$,
 $\sigma_1 \sigma_1 = \sigma_0$,  $\sigma_2 \sigma_1 = i \sigma_3$
and  $\sigma_3 \sigma_1 = -i \sigma_2$, and the fact that the
Pauli matrices are hermitian, we can see that preoperating with
$\sigma_1 \otimes \sigma_1$ does not make any difference to the
action of this channel, and therefore:
\begin{equation}
\cal{E}' = \cal{E}.
\end{equation}
We can also trivially say the same thing if we preoperate with the
identity operation $\sigma_0 \otimes \sigma_0$. Let us now
consider the following `averaging' preoperation:
\begin{equation}
{\cal{F}}(\rho) = {1 \over 2} (\sigma_0 \otimes \sigma_0 \rho
\sigma_0 \otimes \sigma_0 + \sigma_1 \otimes \sigma_1 \rho
\sigma_1 \otimes \sigma_1)
\end{equation}

From the arguments above follows immediately the equality
\begin{equation}
\cal{E} \circ \cal{F} = \cal{E}
\end{equation}
i.e. preoperating on our state with $\cal{F}$ does not affect the
operation of the above channel.  Since by construction $\cal{F}$
corresponds to averaging over the group $\{\sigma_0 \otimes
\sigma_0, \sigma_1 \otimes \sigma_1\}$ we need only to consider
input states that are invariant under it. Let us denote by $R$ the
whole set of 2 qubit density matrices. We are looking for the
explicit form of an input state $\rho \in R$ which minimizes the
output entropy. If we find such an optimal state $\rho_{*}$, then
by the above arguments the input state ${\cal{F}}(\rho_{*})$ will
also give the same output entropy, and will therefore also be
optimal. This means that instead of looking for the optimal state
in $R$, we can instead restrict our search to finding an optimal
state from the restricted set ${\cal{F}}(R)$. Since the optimal
state $\varrho_{*} \in {\cal{F}}(R)$ that minimizes the output
entropy is by construction invariant under the group $\{\sigma_0
\otimes \sigma_0, \sigma_1 \otimes \sigma_1\}$ it can easily be
checked that in the basis
$\{|00\rangle,|01\rangle,|10\rangle,|11\rangle\}$ represented by
the eigenvectors of $\sigma_1 \otimes \sigma_1$ it must take the
form
\begin{equation}
  \varrho_{*} = \left(\begin{array}{cccc}
    a  & 0 & 0 &  c \\
    0 & d & f &  0 \\
    0 &  f^* &  e &  0 \\
     c^* &  0 & 0 & b \end{array}\right)
\end{equation}

From the form of $\rho_{*}$  follows that it is a convex
combination of pure states of the form

\begin{eqnarray}
&& \alpha |00 \rangle + \beta |11 \rangle ~~ \mbox{or} \nonumber
\\ && \alpha |01 \rangle + \beta |10 \rangle\;.
\label{ansatz}
\end{eqnarray}

We will now prove that to minimise the output von Neumann entropy
we can restrict our attention to an input pure state of the form
(\ref{ansatz}). Let us write $ \rho_{*}$ in terms of its pure
state decomposition
\begin{equation}
 \rho_{*} = \sum_{i} p_i |\psi_i \rangle \langle \psi_i |\;.
\end{equation}
Then the action of the channel will give
\begin{equation}
 {\cal{E}}(\rho_{*}) = \sum_{i} p_i
{\cal{E}}(|\psi_i \rangle \langle \psi_i
 |)
\end{equation}
and hence by the concavity of the von Neumann entropy entropy \cite{Nielsen}
we have
\begin{equation}
 S({\cal{E}}(\rho_{*})) \geq
\sum_{i} p_i  S({\cal{E}}(|\psi_i \rangle \langle \psi_i
 |))\;.
\end{equation}
In particular suppose without loss of generality that $|\psi_1
\rangle$ is the pure state in the decomposition of $\rho_{*}$ that
gives the lowest output entropy from all the eigenvectors of
$\rho_{*}$. Then the above equation implies that
\begin{equation}
 S({\cal{E}}(\rho_{*})) \geq S({\cal{E}}(|\psi_1 \rangle \langle
 \psi_1
 |))\;.
\end{equation}
So indeed, as we have assumed that $\rho_{*}$ is already optimal,
this means that this last equation is actually a strict equality,
and hence one of its eigenvectors will also be optimal, namely
\begin{equation}
 S({\cal{E}}(\rho_{*})) = S({\cal{E}}(|\psi_1 \rangle \langle
 \psi_1 |))\;.
\end{equation}
Therefore, we can restrict our attention to finding an input pure state
of the form (\ref{ansatz}).

Let us rewrite without loss of generality the input state (\ref{ansatz}) as

\begin{equation}
|\psi_{\theta,\phi}\rangle
=\cos\theta |00\rangle + e^{i\phi} \sin\theta  |11\rangle\;.
\end{equation}

The corresponding state at the output of the channel takes the form

\begin{eqnarray}
&&{\cal{E}}(|\psi_{\theta,\phi} \rangle \langle
 \psi_{\theta,\phi}|)=\frac{1}{4}\left[ \sigma_0\otimes \sigma_0
+ \eta\cos 2\theta
(\sigma_0\otimes \sigma_1+\sigma_1\otimes\sigma_0) \right.\nonumber\\
&&\left.+ C \sigma_1\otimes\sigma_1
+\mu\sin 2\theta \cos\phi (\sigma_2\otimes\sigma_2-\sigma_3\otimes\sigma_3)
\right.\nonumber\\
&&\left.
+\mu(4p-1)\sin 2\theta \sin\phi (\sigma_2\otimes\sigma_3+\sigma_3\otimes
\sigma_2)\right]\;,
\end{eqnarray}

where $\eta=(4p-1)$ and $C=\mu+(1-\mu)\eta^2$.
As we can easily verify, the above density operator has the following
eigenvalues

\begin{eqnarray}
&&\lambda_{1,2}=\frac{1}{4}(1-C)\nonumber\\
&&\lambda_{3,4}=\frac{1}{4}(1+C)\nonumber\\
&&\pm \frac{1}{2}\sqrt{\eta^2\cos^2 2\theta
+\mu^2\sin^2 2\theta(\cos^2\phi+\eta^2\sin^2 \phi)}\;.
\end{eqnarray}

As we can infer from the above form of the eigenvalues, the input state
corresponding to the minimum entropy is given by $\phi=0$. Moreover,
when $\mu>\eta$, or equivalently $p<(\mu+1)/4$, the input state
with minimum entropy is the maximally entangled state (\ref{ansatz})
with $\theta=\pi/4$. In the other case, when $\mu< 4p-1$, the input state
corresponding to the minimum output entropy is a product state of the form
$|00\rangle$.

The set of optimal 16 states discussed above, that maximizes the mutual information along the channel, reduces
in these cases to a set of four equiprobable input orthogonal states. Therefore, similarly to the case of the
depolarizing channel with memory \cite{MP}, we can identify the onset of a threshold value $\mu_t=4p-1$, above
which the mutual information along the channel is maximized by using equiprobable Bell states. Below the
threshold the use of entanglement does not bring any benefit since the information is optimized by transmitting
product states, such as the set $\{|00\rangle, |01\rangle, |10\rangle, |11\rangle\}$. It is noteworthy that the
resulting channel capacity is a non-differentiable function of the memory parameter $\mu$.

In conclusion, we have studied the performance of Pauli channels
with memory effects for the transmission of classical information,
and we have provided a complete proof that a certain class of
Pauli channels exhibits the onset of a threshold on the degree of
memory. We have shown that below this threshold the two qubit
capacity of the channels is achieved by input product states,
while above it the capacity is achieved by  maximally entangled
input states. This is the first time that entanglement is
rigorously proven to be a precious resource in the transmission of
classical information in the presence of noise. Our results so far
have covered a class of Pauli channels, characterized by a single
noise parameter. However, we have numerical evidence that the
onset of the threshold, and the corresponding enhancement of
information transmission by using entangled states, are features
of most two-qubit Pauli channels with correlated noise \cite{mpv}.

This work has been supported in part by the EC program QUPRODIS
(Contract No. IST-2002-38877) and the US Army Grant No. DAAD
19-02-0161.

%\end{multicols}

\begin{references}
\bibitem{general} The channel that we have chosen is a
probabilistic application of unitary transformations drawn from a
particular group, and we wish to force the input states to be
invariant under one of its subgroups (namely $\{\sigma_0 \otimes
\sigma_0, \sigma_1 \otimes \sigma_1\}$). To ensure this we have
picked the probabilities such that any elements drawn from the
same left coset of the subgroup have identical weight. This is why
the channel that we consider is invariant under pre-operating with
the `averaging over the subgroup' operation, and the same
reasoning also applies to any subgroup/channel combination such
that the weights are constant over each left coset.
\end{references}
\end{document}